\documentclass[useAMS,usenatbib,usegraphicx]{mn2e}

\title[Grand design lifetime]{The lifetime of grand design}
\author[M.R. Merrifield, R.J. Rand and S.E. Meidt]
{M.R.\ Merrifield,$^1$\thanks{E-mail: michael.merrifield@nottingham.ac.uk}
R.J. Rand$^2$ and S.E. Meidt$^2$\\
$^1$School of Physics \& Astronomy, University of Nottingham, 
    University Park, Nottingham, NG7 2RD\\
$^2$Department of Physics and Astronomy, University of New Mexico, 
    800 Yale Boulevard Northeast, Albuquerque, NM 87131, USA
}
\begin{document}

\date{Accepted 2005 November 1.  Received 2005 October 21; in original form 2005
September 17}

\pagerange{\pageref{firstpage}--\pageref{lastpage}} \pubyear{2005}

\maketitle

\label{firstpage}

\begin{abstract}
The lifetime of the structure in grand design spiral galaxies is
observationally ill-determined, but is essentially set by how
accurately the pattern's rotation can be characterized by a single
angular pattern speed.  This paper derives a generalized version of
the Tremaine--Weinberg method for observationally determining pattern
speeds, in which the pattern speed is allowed to vary arbitrarily with
radius.  The departures of the derived pattern speed from a constant
then provides a simple metric of the lifetime of the spiral structure.
Application of this method to CO observations of NGC~1068 reveal that
the pattern speed of the spiral structure in this galaxy varies
rapidly with radius, and that the lifetime of the spiral structure is
correspondingly very short.  If this result turns out to be common in
grand-design spiral galaxies, then these features will have to be
viewed as highly transient phenomena.
\end{abstract}


\section{Introduction}\label{sec:intro}
Spiral arms are some of the most strikingly beautiful features of
galaxies, yet even now we lack a full understanding of their origins.
It is well known that these features cannot simply be interpreted as
long-lived aesthetically-organized ``material spirals'' of
particularly bright stars -- as stressed by \citet{o62}, the strong
differential rotation in disk galaxies would rapidly wind up and
destroy any such arrangement.  It was this ``winding dilemma'' that
motivated \citet{ls66} to build on the earlier intuition of
\citet{l51} and develop a theory by which spirals could persist as
long-lived density waves, thereby apparently solving the problem.
However, ever since this theory was first espoused, questions have
remained.  For example, \citet{t69} argued that such spiral density
waves would tend to dissipate on relatively short timescales and must
be repeatedly re-excited via a swing-amplifier, while \citet{sk91}
presented the interesting alternative ``groove mode'' mechanism for
their initial excitation, in which spiral structure is produced by a
pinching of the disk.  \citet{s00} has also argued that almost all
numerical simulations strongly support the notion that spiral
structure is transient; indeed recent attempts to impose any kind of
long-lived spiral density wave on a numerical simulation only seem to
work if the spiral structure is very tightly wound up \citep{ykg03},
which is not what one finds in many real grand-design spirals.  From
an observational perspective, it has also often been noted that the
most dramatic spiral galaxies like M51 all seem to have interacting
companions, which suggests that such interactions play a role in
continuously exciting an evolving spiral structure.  This anecdotal
impression was borne out in the study by \citet{ee83}, which showed
that grand design spirals are significantly more common in group
environments where such interactions will be more frequent.

There therefore remains the most basic question about the spiral
structure that defines the appearance of so many galaxies: how long
does it persist, or, put more evocatively, how long could
extragalactic travellers stay away and still recognize their own
galaxy when they got home?  This question is of more than purely
navigational interest, as the longevity of spiral structure has
fundamental implications for our understanding of galaxy formation and
evolution.  The interpretation of the Hubble sequence depends strongly
on the duty cycle of the open, grand design spiral structure that
defines the late end of the sequence, while the timescales for
dynamical evolutionary processes such as the heating of the stellar
component by spiral arms \citep{jb90} also clearly depend on how long
these structures last.

The degree to which a pattern persists is dictated primarily by
whether it has a single ``pattern speed,'' $\Omega_p$, the angular
rate at which the pattern rotates.  One of the difficulties in
studying this quantity is that it has no simple relation to the more
directly observable physical velocity of material at different radii
in the galaxy.  The situation is further complicated by the fact that
only one component of this velocity is directly accessible through the
Doppler shift.  However, \citet[hereafter TW]{tw84} elegantly
demonstrated that one can invoke the continuity equation to derive
$\Omega_p$ from the observable material kinematics solely by making
the assumptions that the kinematic tracer is conserved, that it orbits
in a single plane, and that the pattern rotates rigidly, so that
$\Omega_p$ is a single well-defined constant.  Specifically, one can
write the continuity equation as
\begin{equation}
{\partial \over \partial t}\Sigma + {\partial \over \partial x}\Sigma v_x 
+ {\partial \over \partial y}\Sigma v_y = 0,
\label{eq:cont}\end{equation}
where $\Sigma(x,y,t)$ is the surface density of material orbiting in a
plane with the Cartesian coordinates chosen to be aligned with the
galaxy's observed major and minor axes.  TW showed how one can
integrate this equation twice to get rid of the unobservable velocity
component and eliminate the numerically-unstable spatial derivatives,
leaving
\begin{equation}
\Omega_p \int_{-\infty}^{\infty} \Sigma x dx = 
\int_{-\infty}^{\infty} \Sigma v_y dx.
\label{eq:TW}\end{equation}
The terms in both integrals are all measurable ($v_y$ is just the
observable line-of-sight velocity corrected for the galaxy's
inclination).  These integrals can be determined for a range of cuts
at different distances $y$ from the galaxy's major axis.  In
principle, each such cut provides an independent measurement of
$\Omega_p$, but a more robust approach to calculating the pattern
speed involves making a plot of one of these integrals suitable
normalized, $\langle x \rangle = \int \Sigma x dx/\int \Sigma dx$
against the other with the same normalization, $\langle v_y \rangle =
\int \Sigma v_y dx/\int \Sigma dx$: such a plot should yield a
straight line passing through the origin, whose slope is $\Omega_p$\
\citep{mk95}.

The TW technique has now been successfully applied to a number of
barred galaxies, using stars as the tracer population and optical
absorption line spectroscopy to measure their kinematics, with the
interesting result that these objects do have well-defined pattern
speeds and are rapidly rotating \citep{adc03, gkm03, c04}.  However,
analyses of spiral galaxies have generally proved less straightforward
for several reasons.  First, spiral structure is geometrically a good
deal more complicated than a bar.  This more complex structure means
that some parts of the structure may be at an orientation in which the
TW method has no discriminatory power (essentially because $\langle
x\rangle$ and $\langle v_y\rangle$ both become very small).  It is
also the case that some parts of the complex spiral structure are
clearly transient features that take no part in any global pattern
rotation, but can nonetheless produce significant spurious signal in
the TW analysis.  Fortunately, the fact that we can choose the values
of $y$ to which the analysis is to be applied means that we can to a
large extent avoid the regions compromised in this way.  A second
issue is that obscuration and star formation in spiral arms mean that
application of the continuity equation to the observable stellar
component of these systems may not be valid (although may perhaps be
possible using near infrared emission).  The technique has therefore
usually been applied to radio observations of the gaseous components,
either in HI \citep[e.g.][]{w98} or CO \citep[e.g.][]{rw04}, to avoid
problems of obscuration.  Formally, conversion of gas into stars means
that the continuity equation is not applicable to this tracer, either,
but star formation does not have the disproportionate effect on gas
that bright young stars have on the optical emission, so the analysis
is not unduly compromised.

The most interesting problem in the application of the TW method to
spiral galaxies, however, is that even after the above issues have
been dealt with the method still seems to break down because plots of
$\langle x \rangle$ against $\langle v_y \rangle$ do not yield
straight lines for any choice of major axis position angle
[\citet{d03} and \citet{rw04} explore in detail how such plots can be
sensitive to this parameter].  In some cases, the departures from a
line can be attributed to the limited quality of the data, but in
others clear systematic departures from the linear fit are apparent.
By far the simplest interpretation of this failure is that these
spiral arms do not have a single pattern speed, so their structure
must indeed be evolving with time.

In principle, one can use this breakdown of the TW method to constrain
the spatial variation in $\Omega_p$ and hence the lifetime of the
current spiral structure.  However, the original TW method was not
particularly well formulated for such an analysis because the
integrals in equation~(\ref{eq:TW}) combine data from a wide range of
radii, so they produce complicated spatially-averaged estimates of any
varying pattern speed, which have no simple physical interpretation.
\citet{w98} was able to use these values to derive the spatial
variation in $\Omega_p$ under the assumption that it only changes
slowly with position, but such slow variations are by no means
guaranteed.  He also presented a generalization of the technique that
resulted in a tractable integral equation if $\Omega_p$ only varies
with position along the major axis of the galaxy, $x$, but since this
choice of Cartesian coordinate arises simply from the viewing angle
and not any intrinsic property of the galaxy, the pattern speed will
not in reality depend solely on this coordinate.

This paper therefore revisits the problem and derives a generalization
of the TW method in which the pattern speed is allowed to vary
arbitrarily rapidly in the radial direction within the galaxy.  This
is not the only possible way that $\Omega_p$ could vary spatially, but
it is the simplest physically-motivated option.  It explicitly allows
for the possibility that a galaxy may contain a number of distinct
features at different radii, such as bars and spiral arms, each with
their own pattern speeds, and it also permits one to calculate a
simple metric of the lifetime of a galaxy's current pattern.  The
remainder of this paper is laid out as follows.  Section~\ref{sec:TWR}
presents the mathematical generalization of the TW method to a pattern
speed that varies with radius, and Section~\ref{sec:example} gives an
example of its application to the grand design spiral galaxy NGC~1068.
Finally, conclusions are drawn in Section~\ref{sec:conc}.

\section{The Generalized TW Method}\label{sec:TWR}
Even if the pattern speed is allowed to vary with radius within a
galaxy, it is still the case that the pattern at a given radius simply
rotates around in azimuth $\phi$ with time, so one can replace the
temporal derivative in equation~(\ref{eq:cont}) with
\begin{equation}
{\partial \Sigma \over \partial t} 
    = -\Omega_p(r) {\partial \Sigma \over \partial \phi},
\end{equation}
as in the original TW analysis.  Making this substitution in
equation~(\ref{eq:cont}), and integrating over all values of $x$ to
eliminate the unobservable $v_x$ component of velocity, we find
\begin{equation}
\int_{-\infty}^\infty \Omega_p(r) {\partial \Sigma \over \partial \phi}\,dx 
    - \int_{-\infty}^\infty {\partial \over \partial y} \Sigma v_y\,dx = 0.
\end{equation}
Still following TW, we now integrate over $y$ to eliminate a
noise-sensitive spatial derivative, yielding
\begin{equation}
\int_{y'=y}^\infty \int_{x=-\infty}^\infty 
               \Omega_p(r) {\partial \Sigma \over \partial \phi}\,dx\,dy'
    + \int_{-\infty}^\infty \Sigma v_y\,dx = 0.
\end{equation}
Changing variables from Cartesians to polars in the double integral gives
\begin{equation}
\int_{r=y}^\infty \int_{\phi=\arcsin(y/r)}^{\pi - \arcsin(y/r)} 
               \Omega_p(r) {\partial \Sigma \over \partial \phi} r\,dr\,d\phi 
    + \int_{-\infty}^\infty \Sigma v_y\,dx = 0,
\end{equation}
and carrying out the integration with respect to $\phi$ then yields the
final integral equation, 
\begin{equation}
\int_{r=y}^\infty 
 \left\{ 
 \left[\Sigma(x', y)  - \Sigma(-x', y)\right] r \right\} 
                                             \Omega_p(r)\,dr
   = \int_{-\infty}^\infty \Sigma v_y\,dx,\!\!\label{eq:TWR}
\end{equation}
where $x'(r,y)=\sqrt{r^2 - y^2}$. As a check, it is straightforward to
show that equation~(\ref{eq:TWR}) reduces to equation~(\ref{eq:TW}) in
the case where $\Omega_p$ is constant.

These coordinates can be straightforwardly constructed from the
observed coordinates, allowing for the galaxy's inclination: $x =
x_{\rm obs}$, $y = y_{\rm obs}/\cos i$ and $v_y = v_{\rm obs}/\sin i$.
Since the kernel term in curly brackets and the integral on the right
hand side of equation~(\ref{eq:TWR}) are both
observationally-determined quantities, the problem is now reduced to a
Volterra integral equation of the first kind for $\Omega_p(r)$.  Such
equations are straightforward to solve numerically \citep[chapter
18]{nr}: reduction of the integral to a discrete quadrature for
different values of $r=r_i$ and $y = y_j$ turns this problem into a
matrix equation of the form
\begin{equation}
\sum_{r_i > y_j} K(r_i, y_j) \Omega_p(r_i) = f(y_j).\label{eq:TWmat}
\end{equation}
The infinite upper limit on the integral complicates the analysis a
little, but as for the original TW method one finds that the decline
in $\Sigma$ with radius and the lack of strong spiral structure at
large radii mean that the kernel decreases rapidly to zero, so that
the integral has converged at an attainable radius.  The lower limit
on the integral means that the matrix ${\mathbfss K}$ is triangular,
and hence that equation~(\ref{eq:TWmat}) is readily soluble by back
substitution to obtain $\Omega_p(r_i)$.  However, the triangular form
of the matrix does mean that a problem with the data at some radius
will propagate inward to compromise all the derived pattern speeds at
smaller radii, so a degree of care must be taken in evaluating the
range of radii over which pattern speeds can be determined reliably.
Note, though, that in principle one gets two independent measurements
of $\Omega_p(r)$, doubling the chances of finding a region free of
such issues: here we have assumed that $y > 0$, but we could equally
well derive the function from the data at $y < 0$.

As formulated here, $\Omega_p$ can vary arbitrarily with radius.  One
could impose a degree of smoothness by regularizing the solution
through an algorithm that penalizes rapid fluctuations in
$\Omega_p(r)$, or even by finding a suitably-parameterized smooth
function that best matches the two sides of equation~(\ref{eq:TWR}).
However, for this initial test we will avoid imposing any such
prejudice upon the answer, as there is always a danger that a
regularized solution owes its form more to the prior assumption about
the smoothness of the function than the unpredictable astronomical
truth.  This freedom is particularly valuable in studies of pattern
speed, where spatially distinct features can have completely different
pattern speeds, leading to possible discontinuities in $\Omega_p(r)$.

Once the variation in pattern speed with radius has been calculated
from the solution of equation~(\ref{eq:TWmat}), it is simple to
evaluate how the current spiral pattern will evolve into the future.
By following this shearing, different metrics of the lifetime of the
spiral structure can be constructed.  One straightforward
model-independent measure is provided by calculating the dispersion in
the derived values of $\Omega_p(r_i)$, $\sigma(\Omega_p)$; a measure
of the lifetime of the pattern is then just $2\pi/\sigma(\Omega_p)$.
Even simpler, one can just take the largest and smallest values within
some range of radii, and calculate a characteristic wind-up time for
this part of the pattern of $\tau_{\rm wind} = 2\pi/(\Omega_p^{\rm max} -
\Omega_p^{\rm min})$.

\section{Application to NGC~1068}\label{sec:example}
As an initial experiment, we have applied this algorithm to the BIMA
Survey of Nearby Galaxies \citep[BIMA SONG;][]{hetal03} CO
observations of the grand-design Sb galaxy NGC~1068, which offers a
test case well suited to this analysis. Although optically classified
as a normal spiral \citep{RSA}, albeit one with a Seyfert nucleus, the
2MASS near-infrared observations \citep{jccsh03} reveal a clear
central bar, which provides a strong candidate for the driver of the
grand design spiral seen in the CO emission.  \citet{rw04} showed that
NGC~1068's gas content is dominated by molecular material and that
there is not much current star formation, so little of the molecular
phase is being converted to or from other phases and the application
of the continuity equation to this component alone is valid.

A further concern in the application of the continuity equation to
these data is how well the observed CO emission traces the molecular
gas surface density, and hence whether the equation can be applied
directly to the data.  In particular, even within the Milky Way it is
known that the conversion between CO emission and molecular gas
column, conventionally termed $X$, shows a metallicity dependence
\citep{bw95}, while there is also evidence that $X$ varies linearly
with metallicity from galaxy to galaxy \citep{blg02}.  However, as can
be seen from \citet{pvc04}, there is essentially no metallicity
gradient in NGC~1068 over the range of radii probed by the data
analyzed here, so $X$ should be close to constant.  There is also the
possibility that $X$ could vary with other parameters.  For example,
enhanced macroscopic opacity effects, more intense radiation fields,
or stronger deposition of CO on to dust grains could render the values
of $X$ in spiral arms systematically different from those in lower
density regions.  However, recent studies of molecular cloud
populations in nearby galaxies with higher resolution and sensitivity
than previously achieved are not finding any evidence for significant
variations in $X$ with environment \citep{repb03, blij03, lbsb05}.
Physically, one might expect $X$ to vary with temperature and density
as $X \propto T \rho^{-0.5}$ \citep{ys91}; the fact that there is so
little observed variation then implies that in the astrophysical
environment this combination of temperature and density varies very
little in regions where there is a significant amount of molecular
material.  Interestingly, even if one does allow $X$ to vary by a
fairly generous factor of two between the arm and interarm regions, it
makes very little difference to a TW-style analysis \citep{zrm04}.
Clearly, the matter has not yet been finally settled, and any
conclusions will have to carry this caveat, but currently there is no
reason to believe that variations in $X$ should compromise this
analysis.

\begin{figure}
\includegraphics[width=84mm]{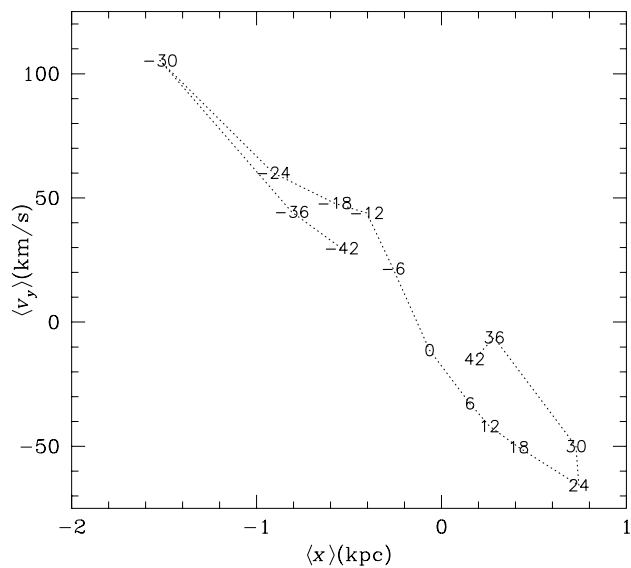}
\caption{Plot of $\langle x\rangle$ versus $\langle v_y\rangle$ for
the BIMA SONG CO observations of NGC~1068.  The numbers indicate the
distance $y_{\rm obs}$ in arcseconds by which each cut is offset from
the galaxy's major axis, and the dotted line joins adjacent cuts.  }
\label{fig:TW}
\end{figure}

What makes NGC~1068 such an interesting target for this analysis is
that the conventional TW analysis produced an interesting result for
this galaxy \citep{rw04}.  As Figure~\ref{fig:TW} reiterates, although
the plot of $\langle x\rangle$ versus $\langle v_y\rangle$ roughly
describes a straight line, it is apparent that there are significant
departures from the expected relation.  It is also clear that these
departures are not the random noise that one would expect if they
simply represented a limited signal-to-noise ratio -- ordering the
points by their distance $y_{\rm obs}$ from the galaxy's major axis
produces a systematic figure-of-eight in this plane.  The deviant
point at $y_{\rm obs} = -30\,{\rm arcsec}$ arises due to a bright spot
in the CO emission that this cut intersects, which is clearly not a
part of the overall spiral structure, so, as discussed above, must be
avoided in the TW analysis.  With the exception of this point,
however, NGC~1068 seems to show evidence for a systematic variation in
pattern speed with position, warranting further investigation through
the analysis developed in this paper.

\begin{figure}
\includegraphics[width=84mm]{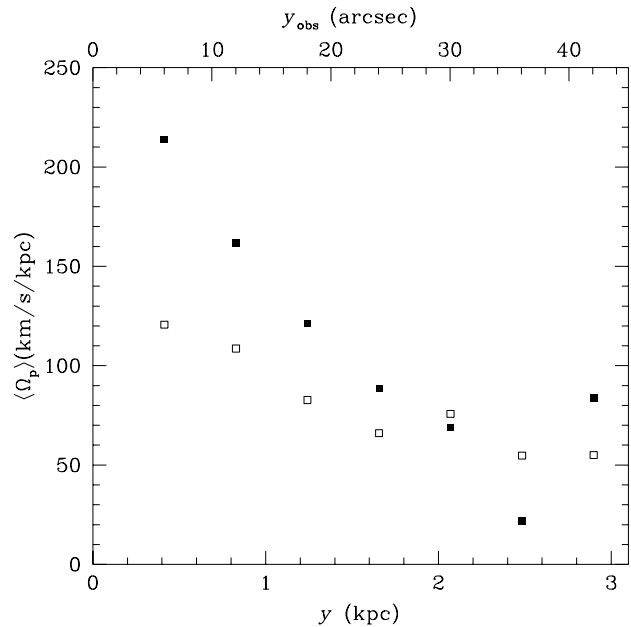}
\caption{Plot of the weighted mean pattern speed as a function of
  distance from NGC~1068's major axis.  Filled symbols are for
  positive $y$ and open symbols show the corresponding values for
  negative $y$.  The scale conversion from arcseconds to kiloparsecs
  has been made by adopting a distance to NGC~1068 of $14.4\,{\rm
  Mpc}$ \citep{bhetal97}.}
\label{fig:TW2}
\end{figure}

A preliminary indication of the variation in pattern speed with radius
can be obtained by calculating $\langle \Omega_p \rangle = \langle v_y
\rangle/\langle x \rangle$ for cuts at different values of $y_{\rm
obs}$.  Figure~\ref{fig:TW2} shows this quantity for both positive and
negative values of $y_{\rm obs}$.  Both sides of the galaxy show a
significant decline in $\langle \Omega_p \rangle$ with radius, which,
as discussed above, implies that NGC~1068 cannot have a constant
pattern speed.  In fact, since the value of $\langle \Omega_p \rangle$
at any given value of $y_{\rm obs}$ is a weighted average of the
pattern speed at all radii $r \ge y$, this plot will if anything tend
to wash out the variation with radius, so we might expect the physical
$\Omega_p(r)$ to be an even more steeply declining function.  It is
also notable that the two sides of the galaxy show a systematic
difference in $\langle \Omega_p \rangle$, which provides
further evidence that the galaxy cannot be described by a global
pattern speed -- interestingly, \citet{w98} found a similar difference
in his analysis of the spiral structure in M81, suggesting that this
phenomenon may be quite common.  It should also be noted that the
deviant point at $y_{\rm obs} = -30\,{\rm arcsec}$ does not show up
dramatically in this plot, as both $\langle x \rangle$ and $\langle
v_y \rangle$ are discrepantly large, so their ratio comes out
more-or-less in line with the other values.  It is only by using the
extra information in Figure~\ref{fig:TW} that the departure from the
grand design structure becomes apparent.

Bearing these points in mind, it is still worth seeking to apply the
generalized TW analysis to see how rapidly the pattern speed would
have to vary with radius to be consistent with the observations.  The
fact that the two halves of the galaxy show somewhat different
properties indicates that the evolution of the pattern must be
somewhat more complex than this radial shearing, but it should
nonetheless give a reasonable first approximation, and an estimate of
the timescale involved for the pattern to evolve significantly.  

\begin{figure}
\includegraphics[width=84mm]{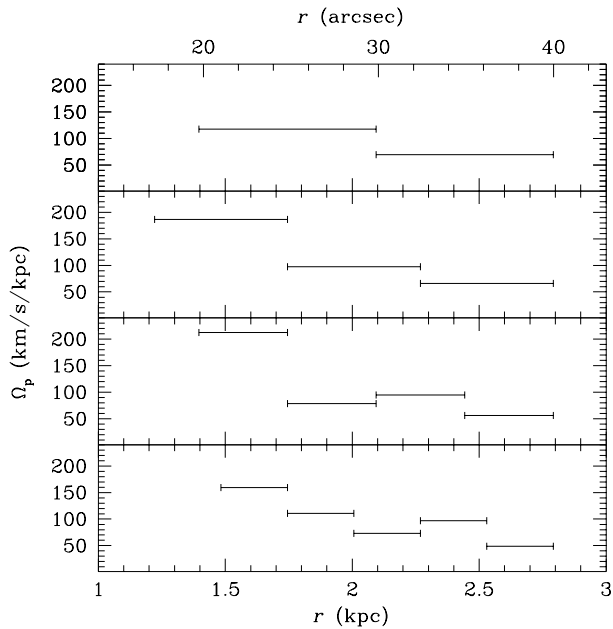}
\caption{Plot of the variation in pattern speed with radius for
  NGC~1068, as derived using the generalized Tremaine-Weinberg method.
  The panels show the results obtained using increasingly
  finely-binned discretizations of the integral
  equation~(\ref{eq:TWR}) to form equation~(\ref{eq:TWmat}).}
\label{fig:TWR}
\end{figure}

To this end, we have solved the matrix equation~(\ref{eq:TWmat}) for
the data at $y > 0$ (thus avoiding the discrepant point at negative
$y$) to estimate $\Omega_p(r)$.  To test the robustness of the result
and estimate the associated errors, Figure~\ref{fig:TWR} shows this
analysis repeated using different radial binnings to discretize the
integral equation~(\ref{eq:TWR}) into the matrix
equation~(\ref{eq:TWmat}).  Even for these ``well-behaved'' data, the
matrix inversion method fails inside $\sim 20\,{\rm arcsec}$.  At
these radii, the galaxy becomes close to axisymmetric, making the
integrals in equation~({\ref{eq:TWR}) small and meaning that noise
starts to dominate.  The further amplification of this noise by the
inversion process then leads to wildly oscillating random values of
$\Omega_p$, a clear sign of numerical instability.  However, at larger
radii the different binnings produce very consistent rather smooth
answers, indicating that the solution to equation~(\ref{eq:TWR}) is
numerically stable in this region even without regularization.  The
variations between different binnings give a good indication of the
uncertainty in $\Omega_p(r)$: although there are variations between
the different binnings, the overall decline in $\Omega_p$ with radius
is robustly reproduced.  As Figure~\ref{fig:TW2} already suggested,
the pattern speed decreases strongly from $\sim 150\,{\rm km/s/kpc}$
at $r \sim 1.5\,{\rm kpc}$ to $\sim 50\,{\rm km/s/kpc}$ at $r \sim
2.5\,{\rm kpc}$.  This analysis fits surprisingly well with an
entirely independent study undertaken by \citet{setgd00}, who looked
at the likely locations for resonances in this galaxy and suggested
that if an inner bar existed at $r \sim 14\,{\rm arcsec}$ then it
should have a pattern speed of $\sim 140\,{\rm km/s/kpc}$, and that an
outer oval-shape distortion which reaches out to $r \sim 120\,{\rm
arcsec}$ may have a pattern speed as low as $\sim 20\,{\rm km/s/kpc}$.
The advantage of the current study is that we are measuring pattern
speeds directly without assuming anything about the resonant radii,
such as presuming that the outer oval structure in this system ends at
its co-rotation radius.  Figure~\ref{fig:TWR} provides direct evidence
that the grand-design spiral structure between the inner bar and outer
oval is not locked to either of their pattern speeds, but rather that
the spiral arm pattern speed varies continuously with radius between
these limiting values.  This strong shearing implies a short lifetime
for the present spiral structure between $r \sim 20\,{\rm arcsec}$ and
$r \sim 40\,{\rm arcsec}$ of only $\tau_{\rm wind} \sim 10^8$ years,
which is directly comparable to (but actually a little shorter than)
the corresponding orbital period at these radii.  This beautiful grand
design spiral would appear to be only a transient feature on the face
of NGC~1068.

\section{Conclusions}\label{sec:conc}
This paper presents a generalization of the Tremaine--Weinberg method
for calculating pattern speeds of galaxies which accommodates the
possibility that the pattern speed may vary with radius.  This
generalization has potential applications to a variety of different
galactic systems.  One could, for example, attempt to determine the
distinct pattern speeds of multiple features within a single galaxy,
such as nested bars-within-bars \citep{cda03}.

Here, however, we have investigated the apparently simpler case of the
grand design spiral structure in the CO emission from NGC~1068.  The
conclusion of this analysis is that this spiral structure does not
have anything close to a single pattern speed, and is winding up on a
timescale marginally shorter than the orbital period, reintroducing
the winding dilemma with a vengeance.  In fact, we know that some of
the other assumptions made in the analysis are violated at some level:
the difference between the kinematic signal at positive and negative
values of $y$ implies that the pattern cannot be evolving with time in
a manner entirely consistent with the simplest azimuthal winding up
considered here.  However, this extra complexity simply means that the
current design must be distorting even more profoundly and rapidly
than we have found.  In the only other comparable study to date,
\citet{w98} tentatively reached a similar conclusion regarding the
grand design spiral structure in M81, based on both differences
between the kinematics of the two sides of the galaxy and the
inconsistency of the results from a TW-style analysis.  If this result
turns out to apply commonly to other galaxies, then intergalactic
travellers would be well advised not to use the morphology of spiral
structure to identify their homes.

\section*{Acknowledgement}
We are very grateful to the referee for helpful comments, which have
significantly improved the presentation of this paper.  MRM is
supported by a PPARC Senior Fellowship, for which he is also most
grateful.

\label{lastpage}

\end{document}